\documentclass[aps,prl,reprint,superscriptaddress,floatfix]{revtex4-2}
\usepackage{amsmath,amssymb,wasysym}
\usepackage{graphicx}
\usepackage{dcolumn}
\usepackage{bm}
\usepackage{braket}
\usepackage{verbatim}
\usepackage{float}
\usepackage{bbold}
\usepackage{color}
\usepackage{xcolor}
\usepackage{relsize}
\usepackage{amsthm}
\usepackage{enumerate}
\usepackage{soul,xcolor}
\usepackage[T1]{fontenc}
\usepackage{adjustbox}
\usepackage[colorlinks=true,urlcolor=blue,urlbordercolor={0 1 1}]{hyperref}

\newcommand{\be}{\begin{equation}}
\newcommand{\ba}{\begin{align}}
\newcommand{\ee}{\end{equation}}
\newcommand{\bea}{\begin{eqnarray}}
\newcommand{\eea}{\end{eqnarray}}
\newcommand{\beq}{\begin{equation}}
\newcommand{\eeq}{\end{equation}}
\newcommand{\beqn}{\begin{eqnarray}}
\newcommand{\eeqn}{\end{eqnarray}}

\newcommand{\rev}[1]{ {\color{red} #1}}
\newcommand{\Ang}{\text{\AA}}
\renewcommand{\hat}[1]{{\widehat #1}}

\usepackage{bm}
\usepackage{array}
\newcolumntype{L}[1]{>{\raggedright\arraybackslash}p{#1}}
\newcolumntype{C}[1]{>{\centering\arraybackslash}p{#1}}
\newcolumntype{R}[1]{>{\raggedleft\arraybackslash}p{#1}}
\usepackage{multirow}
\allowdisplaybreaks

\newcolumntype{C}[1]{>{\centering\arraybackslash}p{#1}}

\usepackage{pifont}
\newcommand{\cmark}{\ding{51}} 
\newcommand{\xmark}{\ding{55}} 

\begin{document}
\title{Pair-density-wave superconductivity and Anderson's theorem in bilayer nickelates}

\author{Hanbit Oh}
\thanks{Corresponding author: \href{mailto:hoh22@jh.edu}{hoh22@jh.edu}}
\affiliation{William H. Miller III Department of Physics and Astronomy, Johns Hopkins University, Baltimore, Maryland, 21218, USA}

\author{Ya-Hui Zhang}
\affiliation{William H. Miller III Department of Physics and Astronomy, Johns Hopkins University, Baltimore, Maryland, 21218, USA}

\date{\today}

\begin{abstract}
The recent experimental observations of high temperature superconductivity in bilayer nickelate have attracted lots of attentions. Previous studies have assumed a  mirror symmetry $\mathcal M$ between the two layers and focused on uniform and clean superconducting states.
Here, we show that breaking this mirror symmetry via an applied displacement field can stabilize a pair-density-wave (PDW) superconductor, which is similar to the Fulde--Ferrell--Larkin--Ovchinnikov (FFLO) state, but at zero magnetic field.
Based on a mean-field analysis of a  model of $d_{x^2-y^2}$ orbital with an effective inter-layer attraction, we demonstrate that the PDW phase is robust over a wide range of displacement field, interlayer hopping strengths, and electron fillings.
Finally, we analyze disorder effects on interlayer superconductivity within the first Born approximation. Based on symmetry considerations, we show that pairing is  weaken by disorders which break the mirror symmetry, even with unbroken time reversal symmetry.  
Our results establish bilayer nickelate as a tunable platform for realizing finite-momentum pairing and for exploring generalized disorder effects.
\end{abstract}

\maketitle

{\it Introduction.---} 
Bilayer nickelates have recently emerged as a promising platform for exploring high-$T_c$ superconductivity, following the report of $80$~K superconductivity in bulk La$_3$Ni$_2$O$_7$~\cite{sun2023signatures,HouJun_2023,PhysRevX.14.011040,Yuanhuiqiu2024}. 
Motivated by their structural similarity to cuprates---yet possessing a distinct electronic configuration with a multi-orbital nature---extensive theoretical~\cite{oh2023type,lu2023interlayer,yang2024strong,qu2023bilayer,Lange2024Pairing,Lange2024Feshbach,lu2023superconductivity,duan2025orbital,zhang2023strong,luo2023bilayer,zhang2023electronic,Zhang2024,sakakibara2023possible,tian2024correlation,qin2023high,yang2023interlayer,gu2025effective,liu2023s,PhysRevB.109.104508,Kakoi2024Pair,Zhu2026Quantum,PhysRevB.110.024514,PhysRevB.110.104507,Luo2024,PhysRevB.109.045154,PhysRevLett.133.096002,cao2023flat,PhysRevB.111.014515,PhysRevB.111.104505,kaneko2025t,ji2025strong,le2025landscape,hu2025electronic,Shao2026Possible,Huang2025Effective,PhysRevB.111.L020504,yang2025strong,oh2024hightemperature,oh2025doping,wang2025originspinstripesbilayer,fan2025minimalbandmodelexperimental} and experimental efforts~\cite{wang2023observation,Wang2024Structure,PhysRevLett.133.146002,Dong2024,Chen2024,Chen2024_chemi,Wang2025,Li_2024,Yang2024,Khasanov2025,Huo2025,10.1093/nsr/nwaf253,bhatt2025resolving,wang2025electronic,sun2025observation,Hao2025,wang2026superconducting} have been dedicated to characterizing this system.
More recently, superconductivity with $T_c \sim 40$~K has also been realized at ambient pressure in strained thin films~\cite{Ko2025,Zhou2025,Liu2025,hao2025superconductivityphasediagramsrdoped}, providing a unique opportunity to probe the system using more accessible measurement techniques, such as angle-resolved photoemission spectroscopy (ARPES) and scanning tunneling microscopy (STM), to elucidate its electronic and magnetic properties. 
On the theoretical side, a wide range of studies have been carried out, proposing various mechanisms for the observed superconductivity. 
One plausible scenario is interlayer $s_\pm$-wave pairing in the $d_{x^2-y^2}$ orbital, driven by strong interlayer superexchange coupling originating from the $d_{z^2}$ orbital via Hund's coupling~\cite{oh2023type,lu2023interlayer,yang2024strong,qu2023bilayer,Lange2024Pairing,Lange2024Feshbach,lu2023superconductivity,duan2025orbital,zhang2023strong}. 
However, most existing studies assume mirror symmetry between the two layers, which guarantees the nesting condition necessary for interlayer pairing. 
The regime where mirror symmetry is significantly broken remains rarely unexplored~\cite{Liu2024Pairdensitywave,Han2024QuantumGeometricNesting}.

\begin{figure}[b]
    \centering
\includegraphics[width=.95\linewidth]{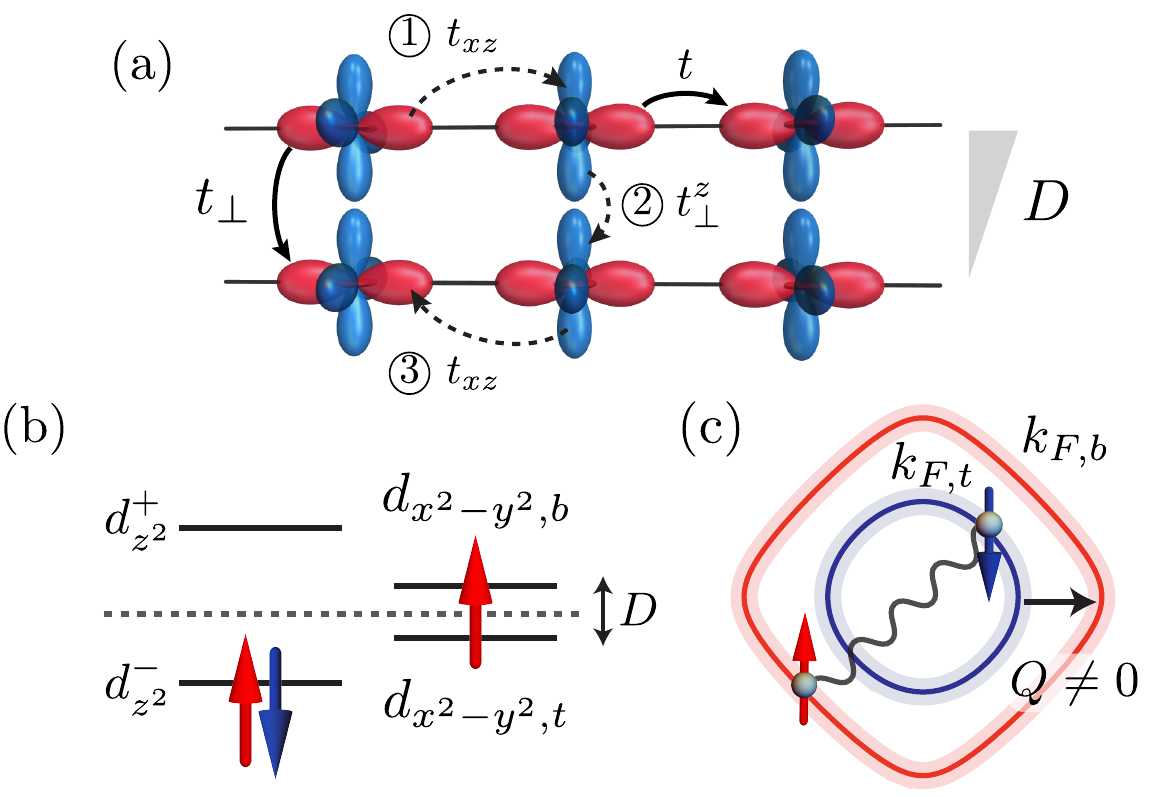}
\caption{(a) Schematic illustration of a bilayer nickelate. Each layer forms a square lattice hosting the $d_{x^2-y^2}$ and $d_{z^2}$ orbitals. In the effective one-orbital description of in $d_{x^2-y^2}$ orbital, we only keep inplane hopping $t$, and effective interlayer hopping $t_\perp$, generated as the third-order process in the underlying two-orbital model (\textcircled{1},\textcircled{2},\textcircled{3} with dashed arrows). 
An external displacement field $D$ breaks mirror symmetry between the two layers. 
(b) The $d_{z^2}$ orbital is nearly half-filled, while the $d_{x^2-y^2}$ orbital is close to quarter filling. 
The energy levels are shown schematically; in reality, the $d_{x^2-y^2}$ orbital forms a dispersive band with average filling $n=1-x$, where $x$ denotes hole doping. 
(c) Fermi-surface mismatch between the two layers leads to Cooper pairing with finite momentum $\delta k_F = k_{F,t} - k_{F,b}$. 
For finite $t_\perp$, the two layers are hybridized and $Q$ is generally defined in the band basis.
    }
    \label{fig:1}
\end{figure}

Bilayer structures introduce internal degrees of freedom that, when symmetry constraints are relaxed, can host unconventional states such as the pair-density-wave (PDW) superconductor~\cite{Agterberg2020Physics}.
A typical realization of a PDW is the Fulde--Ferrell--Larkin--Ovchinnikov (FFLO) state, which relies on Zeeman splitting and explicitly breaks time-reversal symmetry~\cite{fulde1964superconductivity,larkin1964nonuniform}. 
Identifying alternative mechanisms for time-reversal-symmetric PDW states remains a significant challenge, and the search for candidate materials is ongoing. 
In this work, we propose that bilayer nickelates provide a promising platform to realize PDW states by breaking mirror symmetry but at zero magnetic field. 
Specifically, under the strong interlayer pairing scenario, the resulting superconducting state can acquire a finite center-of-mass momentum upon the application of an external electric field. 
The momentum of the PDW is determined by the Fermi-surface mismatch between the two layers. 
Using a mean-field analysis of an effective single-orbital model, we demonstrate the existence of a robust PDW state over a wide range of the phase space.
Although direct evidence for spontaneous
mirror-symmetry breaking in bulk bilayer nickelates is not yet established,
mirror symmetry can be explicitly broken in thin-film or heterostructure
settings, for example by electrostatic gating, polar substrates, or local layer-asymmetric defects.
These settings provide experimentally feasible platforms to test whether
mirror-symmetry breaking can promote finite-momentum interlayer pairing.

In addition, we investigate the interplay between interlayer pairing and disorder in bilayer nickelates. 
Anderson's theorem~\cite{anderson1959theory} states that conventional $s$-wave superconductivity is robust against disorder that preserves time-reversal symmetry. 
This principle has been generalized to multiband and odd-parity superconductors in several studies~\cite{michaeli2012spin,hoyer2015pair,dodaro2018generalization,andersen2020generalized}. 
Here, we focus on interlayer spin-singlet pairing, where the pairing partners are related by the composite $\mathcal{M}\mathcal{T}$ symmetry, combining mirror reflection $\mathcal{M}$ and time-reversal symmetry $\mathcal{T}$. 
By employing the superconducting fitness formalism~\cite{Ramires2018Tailoring,andersen2020generalized} within the conventional Born approximation~\cite{abrikosov1961problem,bruus2004many}, we show that Anderson's theorem for interlayer pairing can be generalized to the $\mathcal{M}\mathcal{T}$-preserving channel. 
Moreover, we examine the effects of finite interlayer hopping $t_\perp$. 
We find that disorder breaking the mirror symmetry $\mathcal{M}$ consistently suppresses interlayer pairing, regardless of the presence of $t_\perp$.

\textit{Effective one-orbital model.---}
The bilayer nickelate can be described by the two-orbital model on square lattice as illustrated in Fig.~\ref{fig:1}(a)~\cite{luo2023bilayer} (see Supplemental Material (SM)~\cite{SM}). We label $d_{x^2-y^2}$ and $d_{z^2}$ orbitals as $d_1$ and $d_2$. Electron filling is $n=2-x$ per site (summed over spin), with $x\approx0.5$ relevant to experiments, corresponding to orbital fillings $n_1 \approx 0.5$ and $n_2 \approx 1$. 
The resulting band structure hosts $\alpha$, $\beta$, and $\gamma$ Fermi pockets, as depicted in Fig.~\ref{fig:2}(d)~\cite{luo2023bilayer}. It has been shown that $\alpha$ and $\beta$ pockets exhibit mixed orbital character, while the $\gamma$ pocket is mainly from the $d_{2}$ orbital.  
Experimentally, the existence of the $\gamma$ pocket remains controversial:
while some ARPES studies report its presence in thin-film samples~\cite{Li_2025},
other measurements do \textit{not} observe it~\cite{wang2025electronic,sun2025observation}.


We now construct an effective one-orbital model to describe the low-energy physics more efficiently. 
This effective one-orbital description is motivated by the strong-correlation regime with large Hubbard $U$, where the nearly half-filled $d_2$ orbital is
expected to be Mott localized while the $d_1$ orbital remains itinerant. As
argued in Refs.~\cite{oh2023type,lu2023interlayer}, we keep the $d_1$
orbital as the low-energy degree of freedom and integrate out the $d_2$ sector, whose interlayer exchange is transmitted to the $d_1$ carriers through Hund's
coupling in later.
 Hence we focus on the $\alpha$ and $\beta$ Fermi pockets and consider an effective single-$d_1$ orbital model,
\begin{eqnarray}
    H_{0}^\mathrm{eff}&=&
    \sum_{k,l,\sigma}
    \xi(\bm{k})\, n_{k,l,\sigma}
    +\sum_{k,\sigma}
    \gamma(\bm{k})\, c_{ k,t,\sigma}^\dagger
    c_{k,b,\sigma}+ \text{H.c.},
    \label{eq:one-orbital}
\end{eqnarray}
with
\begin{eqnarray}
\xi(\bm{k})&=&-2t_x(\cos k_x + \cos k_y)-\mu,
\end{eqnarray}
and
\begin{eqnarray}
\gamma(\bm{k})&=&-t_\perp (\cos k_x -\cos k_y)^2.
\end{eqnarray}
Here, $c_{k,l,\sigma}=d_{1,k,l,\sigma}$ denotes the electron operator for the $d_{1}$ orbital on layer $l$. The first term describes in-plane nearest-neighbor hopping within each layer. 
The second term represents an effective interlayer hopping in the
$d_{x^2-y^2}$ sector. Since the direct interlayer hopping between
$d_{x^2-y^2}$ orbitals is strongly suppressed, with very small
$t_\perp^x$~\cite{luo2023bilayer}, the dominant contribution arises from the
virtual process
$d_{x^2-y^2,t}\rightarrow d_{z^2,t}\rightarrow d_{z^2,b}\rightarrow
d_{x^2-y^2,b}$, involving the inter-orbital hopping $t_{xz}$ and the direct
interlayer hopping of the $d_2$ orbital [Fig.~\ref{fig:1}(a)]. Since each
$d_{x^2-y^2}$-$d_{z^2}$ hybridization carries the factor
$\cos k_x-\cos k_y$, perturbation theory gives the form factor
$(\cos k_x-\cos k_y)^2$ in $\gamma(\bm{k})$ (see SM Sec.~IB~\cite{SM}).
For simplicity, we drop the subscript $x$ and use the notation $t=t_x$ and $t_\perp=t_x^\perp$, and throughout this paper we set $t=1$. The chemical potential $\mu$ is chosen to fix the electron density of the $d_1$ orbital to $n_1=1-x$, where $x$ denotes the hole doping. We mainly focus on $x=0.5$, corresponding to the $d^{7.5}$ electron configuration of the material.
In Fig.~\ref{fig:2}, we plot the band dispersion of Eq.~\eqref{eq:one-orbital} for two representative values, $t_\perp=0.2$ and $0.5$. For sufficiently large $t_\perp$, the interlayer hybridization induces a band splitting along the $\Gamma$--$X$--$M$ line, resulting in an electron pocket near $\Gamma=(0,0)$ and a hole pocket near $M=(\pi,\pi)$. These features reproduce the $\alpha$ and $\beta$ pockets of the original two-orbital model.

\begin{figure}
\centering\includegraphics[width=.9\linewidth]{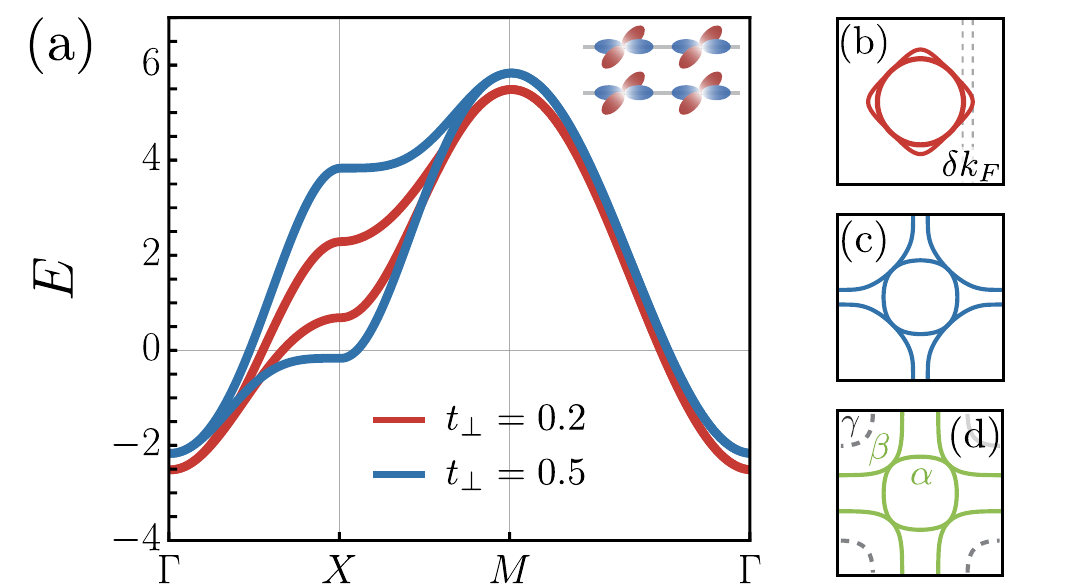}
\caption{(a) Band dispersion of the effective one-orbital model in Eq.~\ref{eq:one-orbital}. 
We set $t=1$ and choose $t_\perp=0.2$ and $t_\perp=0.5$.
The chemical potential is set to $\mu=-1.487$ and $\mu=-1.833$, respectively, to fix the average electron density of the $d_1$ orbital to $n_1=1-x$ with $x=0.5$. 
(b,c) Fermi surfaces for $t_\perp=0.2$ (red) and $t_\perp=0.5$ (blue). 
The interlayer hopping $t_\perp$ splits the two bands and induces a Lifshitz transition. For small $t_\perp$, two electron pockets appear near $\Gamma$, while increasing $t_\perp$ drives one of the electron pockets into a hole pocket at $M$.
(d) For comparison, we show the Fermi surface of the two-orbital model in Eq.~(S1), hosting the $\alpha$, $\beta$, and $\gamma$ pockets.
We stress that the $\alpha$ and $\beta$ pockets touch along the diagonal
direction, which is naturally explained by the vanishing of the form factor
$(\cos k_x-\cos k_y)^2$ on this line.}
    \label{fig:2}
\end{figure}

 We further introduce an external electric displacement field that breaks mirror symmetry between the two layers, which is the central tuning parameter of our study. The resulting total noninteracting Hamiltonian becomes
\begin{eqnarray}
H_{\mathrm{eff}}&=&H^{\mathrm{eff}}_{0}+ 
    D\sum_{k}
    \left(n_{t,k}-n_{b,k}\right),
\end{eqnarray}
where $D$ denotes the strength of the displacement field. 
In the following, we will use this effective model to analyze superconducting instabilities.

\textit{Mean-field theory.---}
We now turn to interaction effects. We consider a phenomenological attractive interaction between the two layers, analogous to the effective interaction in conventional Bardeen--Cooper--Schrieffer (BCS) theory. Specifically, we assume an interlayer interaction of the form,
\begin{eqnarray}
 H_\mathrm{int} &=& 
J_\perp^\mathrm{eff}\sum_{i} \left[ \bm{S}_{t,i}\cdot \bm{S}_{b,i} -\frac{1}{4} n_{t,i} n_{b,i} \right].
\end{eqnarray}
While our aim is to demonstrate the qualitative possibility of a PDW state, a more quantitative description would require a strong-coupling approach based on detailed microscopic models, which we leave for future work.
Microscopically, $J_\perp^{\rm eff}$ should be viewed as an effective interaction inherited from the strong interlayer superexchange of $d_{2}$ sector and transmitted to the itinerant $d_{1}$ carriers through Hund's coupling~\cite{oh2023type,lu2023interlayer}.
This is an effective attraction and a more rigorous analysis would need to incorporate the local moments from the $d_{2}$ orbital and interlayer repulsion~\cite{oh2024hightemperature}. Since we expect our qualitative results to depend primarily on the pairing symmetry, we employ this simplified model for illustrative purposes.

We perform a standard mean-field analysis using an ansatz for the interlayer pairing order parameter,
\begin{eqnarray}
    \langle c_{t,i \uparrow}c_{b,i \downarrow}\rangle = \Delta_{\bm{Q}} \exp(i \bm{R}_i \cdot \bm{Q}),
\end{eqnarray}
where $\bm{Q}$ denotes the center-of-mass momentum of the Cooper pair. The resulting mean-field Hamiltonian is given by,
\begin{eqnarray}
    H_{\mathrm{MF}} = H_{\mathrm{eff}} -
    J_\perp^\mathrm{eff}
    \sum_{\bm{k}} \Delta_{\bm{Q}}\, c_{t,\bm{k}^+}^\dagger (i\sigma_y)\, c_{b, -\bm{k}^-}^\dagger + \text{H.c.},
\end{eqnarray}
where $\bm{k}^{\pm} = \bm{k} \pm \bm{Q}/2$, such that $(\bm{k}^+, -\bm{k}^-)$ forms a Cooper pair carrying a total momentum $\bm{Q}$. The chemical potential $\mu$ is adjusted to fix the total electron density at $n=1-x$. For each $\bm{Q}$, the pairing amplitude $\Delta_{\bm{Q}}$ is determined self-consistently from $H_{\mathrm{MF}}$. To identify the stable phase, we compute the mean-field free energy and compare it across different values of $\bm{Q}$. Since the system lacks perfect nesting and the Fermi-momentum mismatch $|k_{F,+} - k_{F,-}|$ is not uniquely defined, we perform our analysis with $\bm{Q}=(Q,0)$ along the $x$-direction, scanning all possible $Q \in [0, \pi]$. This mean-field approach allows us to determine the energetically favored state as a function of the system parameters.

\begin{figure}[tb]
    \centering
    \includegraphics[width=\linewidth]{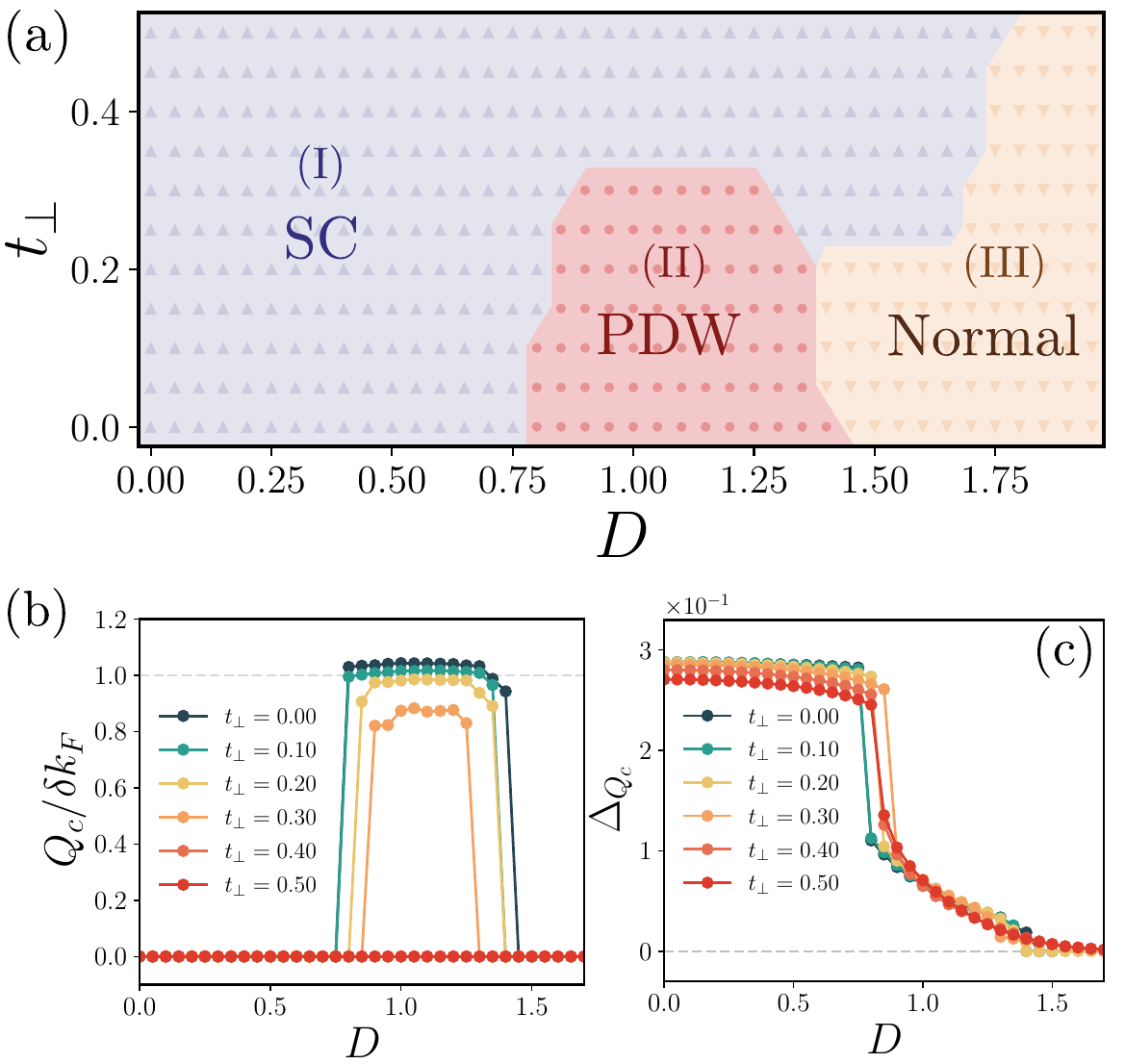}
    \caption{(a) Zero-temperature mean-field phase diagram in the $(D, t_\perp)$ plane. We fix $J_\perp^{\rm eff}= 2$ and $x = 1/2$. As $D$ increases, the system undergoes consecutive transitions from a uniform superconducting (SC) phase to a pair-density-wave (PDW) phase, and finally to a normal state. (b, c) Mean-field solutions that minimize the free energy. For small $t_\perp$, the optimal pairing momentum $Q$ is close to $\delta k_F$, the Fermi-momentum difference along the $x$-direction as illustrated in Fig.~\ref{fig:2}(b), while it deviates from this value as $t_\perp$ increases. The transition between the uniform SC and PDW phases is of first order.}
    \label{fig:3}
\end{figure}

\textit{Pair-density-wave.---}
Our mean-field analysis reveals that breaking mirror symmetry via an external electric displacement field stabilizes a PDW state, similar to the Fulde--Ferrell (FF) phase. The primary results are summarized in Fig.~\ref{fig:3}.

Fig.~\ref{fig:3}(a) presents the zero-temperature phase diagram in the $(D, t_\perp)$ plane at $x=0.5$ and $J_\perp^{\rm eff}=2$, which corresponds to the $d^{7.5}$ electron configuration. Three distinct phases are identified: (I) a uniform superconducting (SC) phase where $\Delta_0 \neq 0$, (II) a pair-density-wave (PDW) phase characterized by $\Delta_{\bm{Q}} \neq 0$ with $\bm{Q} \neq 0$, and (III) a normal metallic phase with $\Delta_0 = \Delta_{\bm{Q}} = 0$ for all $\bm{Q}$. For a fixed $t_\perp$, increasing the displacement field $D$ drives a transition from the uniform SC phase to the PDW phase, with the PDW phase appearing in the intermediate regime. At sufficiently large $D$, strong mirror-symmetry breaking induces a significant energy splitting between the layers, which suppresses interlayer pairing and stabilizes a normal metallic state, even at zero temperature.

Increasing $t_\perp$ tends to destabilize the PDW phase; in the large-$t_\perp$ limit, the system favors a uniform $s^{\pm}$ pairing state instead. Fig.~\ref{fig:3}(b) shows the optimal ordering wave vector $\bm{Q}_c$. For small $t_\perp$, $\bm{Q}_c$ is close to $\delta k_F$, but it deviates as $t_\perp$ increases. This deviation occurs because $\delta k_F$ is defined in the bonding--antibonding band basis, $\delta k_F = k_{F,+} - k_{F,-}$, whereas the pairing occurs between electrons on different layers in the real-space basis. 
Fig.~\ref{fig:3}(c) demonstrates that the transition between the uniform SC and PDW phases is first-order, while the transitions between the PDW and normal phases, or between the SC and normal phases, are continuous. 
We mainly present the representative case with
$J_\perp^{\rm eff}=2$. From the microscopic estimate based on DFT parameters,
one obtains $J_\perp^{\rm eff}\sim 4(t_\perp^z)^2/U \sim 0.65$--$0.8t$~\cite{luo2023bilayer}.
To check the robustness of the PDW phase in this regime, we further perform a calculation at $J_\perp^{\rm eff}=0.75$ (see
Fig.~\ref{fig:4}(a)) and demonstrate that while the PDW region is reduced and the critical
displacement field is shifted to a smaller value $D_c\sim 0.5 t$, the qualitative phase structure
remains unchanged.

To isolate the role of Fermi-surface shape and nesting effect, we also compare with a hypothetical case in which the form factor is removed.
Although the form factor in the interlayer hopping $t_\perp$ is motivated by the specific band structure of bilayer nickelates, we find that it does not qualitatively alter the phase diagram and PDW is more stabilized in case of without, due to better nesting condition (See Fig.~\textcolor{red}{S1}(b,c)). 
Note that in the absence of the form factor, the band splitting becomes nearly uniform along the Fermi surface, allowing a larger portion of the two Fermi surfaces to be connected by a single momentum vector and thereby improving the nesting
condition.
Consequently, similar PDW phases could be realized more generally in other bilayer systems subject to an external electric field. Finally, Fig.~\ref{fig:4}(b) examines the dependence on electron filling, demonstrating that the PDW phase remains robust against doping.

\textit{Generalized Anderson theorem.---}
We have shown that a mirror-symmetry breaking term, such as the displacement field $D$, significantly alters the phase diagram. We now turn our attention to disorder that breaks mirror reflection symmetry only locally. For conventional $s$-wave superconductors, Anderson's theorem states that the pairing is stable against disorder as long as time-reversal symmetry $\mathcal{T}$ remains unbroken. For our interlayer-paired superconductor, however, we demonstrate that disorder breaking mirror symmetry can suppress pairing even when time-reversal symmetry is preserved.

We consider a general spin-independent disorder potential defined in the two-layer basis,
\begin{eqnarray}
    V_{\mathrm{dis}} &=& \sum_{i=0,1,2,3} V_i \, \tau_i \otimes \sigma_0, \label{eq:disorder_pot}
\end{eqnarray}
where $\tau_i$ and $\sigma_i$ are Pauli matrices acting in layer and spin space, respectively. The symmetry properties of the disorder channels are determined by the antiunitary time-reversal operator $\mathcal{T}=i\sigma_y K$, where $K$ denotes complex conjugation, and the unitary mirror-reflection operator $\mathcal{M}=\tau_x$. Accordingly, the four disorder channels are classified as, (i) mirror-symmetric and time-reversal-symmetric ($V_0$), (ii) mirror-symmetric and interlayer time-reversal-symmetric ($V_1$), (iii) mirror-antisymmetric and time-reversal-antisymmetric ($V_2$), and (iv) mirror-antisymmetric and time-reversal-symmetric ($V_3$).

\begin{figure}[tb]
    \centering
    \includegraphics[width=\linewidth]{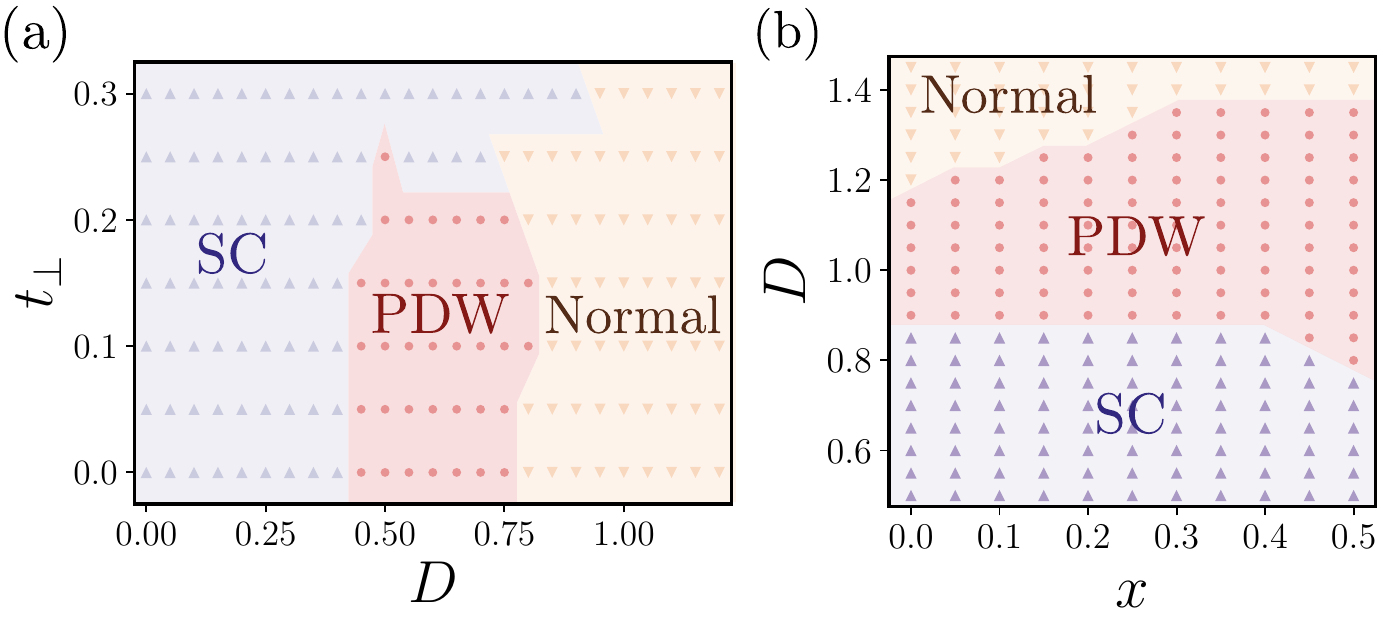}
    \vspace{-10pt}
\caption{(a) Zero-temperature mean-field phase diagram at $J_\perp^{\rm eff}=0.75$ and $x=1/2$. Compared with Fig.~\ref{fig:3}, the result at
$J_\perp^{\rm eff}=2$, the PDW region is reduced and the critical
displacement field is shifted to a smaller value, $D_c\sim 0.5$. Nevertheless, the qualitative phase structure remains unchanged.
(b) Filling dependence of the zero-temperature mean-field phase diagram, where we fix $t_\perp = 0.1$ and $J_\perp^{\rm eff} = 2$. The PDW phase appears over a wide range of filling $x$.}
    \label{fig:4}
\end{figure}

We treat disorder perturbatively within the first-order Born approximation. We first consider the case $t_\perp = 0$, for which the two bands are degenerate. In this limit, following Ref.~\cite{andersen2020generalized}, the disorder-induced suppression of the critical temperature $T_c$ can be expressed in terms of the superconducting fitness function $F_c$~\cite{Ramires2018Tailoring},
\begin{eqnarray}
   \log \left( \frac{T_c}{T_{c,0}} \right) &=& -\frac{\pi}{4} \frac{\alpha_{\mathrm{dis}}}{T_{c,0}} \Gamma, \label{eq:Tc_impurity}
\end{eqnarray}
where $T_c$ is linearly suppressed by the effective scattering frequency, $\Gamma\equiv \pi N(0)n_{\mathrm{imp}} V^2$ is used.  
$T_{c,0}$ denotes the critical temperature in the absence of disorder and $N(0)$ is the density of states at the Fermi energy. The dimensionless depairing coefficient $\alpha_{\mathrm{dis}}$ is given by,
\begin{eqnarray}
    \alpha_{\mathrm{dis}} &=& \frac{1}{4}\mathrm{Tr} \left( \tilde F_c^\dagger \  \tilde F_c \right) , \label{eq:alpha_dis}
\end{eqnarray}
with the fitness function defined as,
\begin{eqnarray}
    \tilde F_c &=& \hat V_{\mathrm{dis}} \hat \Delta - \hat \Delta \hat V_{\mathrm{dis}}^*. \label{eq:fitness}
\end{eqnarray}
It is important to note that $\hat{V}_\mathrm{dis}$, $\hat{\Delta}$ denotes normalized quantity divided by the disorder strength and the superconducting gap. The depairing coefficient $\alpha_{\mathrm{dis}}$ quantifies the robustness of the superconducting state, where a larger $\alpha_{\mathrm{dis}}$ corresponds to a more fragile state, while $\alpha_{\mathrm{dis}} = 0$ indicates complete robustness in the weak-disorder limit.

\begin{table}[t]
    \centering
    \begin{tabular}{c|c|c|c}
    \hline
    \hline
    Disorder & $\mathcal{M}$ & $\mathcal{T}$ & Stab. of $\Delta_\perp$ \\ 
    \hline 
    $V_0$ & $+$ & $+$ & \cmark  \\
    $V_1$ & $+$ & $+$ & \cmark  \\
    $V_2$ & $-$ & $-$ & \cmark  \\
    $V_3$ & $-$ & $+$ & \xmark \\
    \hline
    \hline
    \end{tabular}
       \caption{
    Summary of the generalized Anderson theorem for interlayer spin-singlet pairing, 
    $\Delta_\perp = \tau_x \otimes i\sigma_y$. 
    The second and third columns list the symmetry properties under mirror reflection $\mathcal{M}$ and time-reversal symmetry $\mathcal{T}$ for the four disorder potentials. 
    The fourth column indicate the stability of interlayer pairing for both cases without ($t_\perp = 0$) and with ($t_\perp \neq 0$) interlayer hopping, determined by the depairing coefficient $\alpha_\mathrm{dis}$. 
    A \cmark $\ $ denotes $\alpha_{\mathrm{dis}} = 0$ (stable SC), while a \xmark $\ $ denotes $\alpha_{\mathrm{dis}} \neq 0$ (fragile SC).
    }
    \label{table:1}
\end{table}

For $t_\perp = 0$ and interlayer spin-singlet pairing, $\Delta_\perp = \tau_x \otimes i\sigma_y$, we derive that $F_c = 0$ when the disorder potential possesses $\mathcal{MT}$ symmetry. This implies that interlayer spin-singlet superconductivity is stable under $\mathcal{M}\mathcal{T}$-preserving disorder scattering. Accordingly, Anderson's theorem for interlayer pairing should be modified to require $\mathcal{M}\mathcal{T}$ symmetry preservation. Applying the same analysis to intralayer pairing, $\Delta_\parallel = \tau_0 \otimes i\sigma_y$, we recover the conventional Anderson theorem, where the non-pair-breaking condition is simply the preservation of $\mathcal{T}$. The results are summarized in Table~\ref{table:1}.

Another important question is how this generalized Anderson theorem is modified in the presence of finite interlayer hopping $t_\perp$. To address this, we extend our analysis of $\alpha_{\mathrm{dis}}$ to the $t_\perp \neq 0$ case. 
Our derivation reveals that the expression for $\alpha_{\mathrm{dis}}$ (Eq.~\ref{eq:alpha_dis}) remains invariant even under the band splitting induced by $t_{\perp}$. Hence underlying symmetry plays a more important role in determining robustness under weak disorder than the specific microscopic details of the Hamiltonian (See SM for details~\cite{SM}).

Consequently, the generalized Anderson theorem established here remains applicable in realistic experimental settings where disorder is typically time-reversal symmetric. Notably, for mirror-symmetry-breaking disorder ($V_3$), $\alpha_\mathrm{dis}$ remains non-zero for both $t_\perp=0$ and $t_\perp \neq 0$, even though the clean-limit critical temperature $T_{c,0}$ is rescaled by the modified density of states (DOS).
Asymmetric apical oxygen vacancy is a natural point-defect realization of the $\tau_3$ disorder channel.
In real materials, individual defects are likely to break mirror symmetry locally. Therefore, $T_c$ in bilayer nickelates is expected to be sensitive to disorder, and  higher $T_c$ values can be achieved in cleaner samples.


\textit{Discussion and Conclusion.---}
In this work, we have shown that breaking mirror symmetry via a displacement field in bilayer nickelates provides a natural tuning knob for realizing a pair-density-wave (PDW) state. 
The PDW state achieved here is analogous to the FFLO state in the sense that both arise from Fermi-surface mismatch. However, unlike the FFLO state, here the mismatch is induced by a displacement
field that breaks mirror symmetry while preserving time-reversal symmetry. This provides a potentially more favorable route to finite-momentum pairing, since
it does not rely on a magnetic field.
Within a mean-field framework, we demonstrated that the PDW phase is robust over a wide region of the parameter space.
Beyond the weak-coupling mean-field analysis considered here, our subsequent work extended this mechanism to a strong-coupling one-dimensional regime using density-matrix renormalization group (DMRG) and bosonization analyses~\cite{oh2026incommensurate}.
We demonstrated the stabilization of PDW correlations
and further identified the instability toward charge-$4e$ superconductivity via $t_\perp$ term.

For the microscopically estimated value $J_\perp^{\rm eff}=0.75$, the onset of the PDW phase occurs around $D/t\sim 0.5$ (see Fig.~S2), corresponding to an electric field $E_{\rm PDW}\sim 0.06$--$0.08\mathrm{V}/\Ang$ for $d\sim 3$--$4\Ang$. This is comparable to the experimentally relevant high-field scale $E\sim 0.1\mathrm{V}/\Ang$, suggesting that the PDW regime may be accessible in realistic thin-film settings.
We also comment on two limitations of this work. First, while our work is based on a one-orbital model for the $d_{x^2-y^2}$ sector, a full two-orbital
treatment including itinerant $d_{z^2}$ electrons and a possible $\gamma$ pocket is an interesting direction for future work. Second, this work focuses on the regime where the interlayer attraction $J_\perp^{\rm eff}$ dominates over the in-plane exchange $J^{\rm eff}$. When these two energy scales are comparable, an intralayerr $d$-wave pairing channel can compete with the interlayer $s$-wave channel and may also compete with the displacement-field-induced PDW state.
Finally, through an analysis of disorder effects using the superconducting fitness formalism, we generalized Anderson’s theorem to interlayer spin-singlet pairing, showing that mirror-breaking disorder suppresses superconductivity. 
Our results establish bilayer nickelates as a tunable platform for engineering finite-momentum superconductivity and exploring symmetry-protected robustness against disorder.

\textit{Note added.---} 
During the finalization of this manuscript, we became aware of a preprint~\cite{fan2025minimalbandmodelexperimental}, which also studies a displacement-field-induced Fulde--Ferrell state in a one-orbital description under strong-coupling regime. 
However, our work has a complementary focus beyond the clean-limit PDW phase
diagram. We analyze disorder effects and derive a generalized Anderson theorem
for interlayer pairing, which were not addressed in Ref.~\cite{fan2025minimalbandmodelexperimental}. Conversely,
Ref.~\cite{fan2025minimalbandmodelexperimental} further studies the weak-coupling scenario and the
displacement-field-driven transition between $s_\pm$ and $d$-wave pairing.

\textit{Acknowledgements.---}
H.O. and Y.-H.Z. are supported by a startup fund from Johns Hopkins University and the Alfred P. Sloan Foundation through a Sloan Research Fellowship (Y.-H.Z.).

%

\onecolumngrid
\newpage
\clearpage
\setcounter{equation}{0}
\setcounter{figure}{0}
\setcounter{table}{0}
\setcounter{page}{1}
\setcounter{section}{0}

\maketitle 
\makeatletter
\renewcommand{\theequation}{S\arabic{equation}}
\renewcommand{\thefigure}{S\arabic{figure}}
\renewcommand{\thetable}{S\arabic{table}}
\renewcommand{\thesection}{S\arabic{section}}

\appendix
\onecolumngrid

\begin{center}
\vspace{10pt}
\textbf{\large Supplemental Material for ``Pair-density-wave superconductivity and Anderson's theorem in bilayer nickelates''}
\end{center} 
\begin{center}
{Hanbit Oh$^{1,\textcolor{red}{*}}$, and Ya-Hui Zhang$^{1}$}\\
\emph{$^{1}$ William H. Miller III Department of Physics and Astronomy, \\
Johns Hopkins University, Baltimore, Maryland, 21218, USA}\\

\vspace{5pt}
\end{center}
\tableofcontents

\section{I. From two-orbital model to one-orbital model}
\label{sec:s1}
\subsection{A. Two-orbital model}
We start from a two-orbital tight-binding model on a bilayer square lattice, described by the following Hamiltonian:
\begin{eqnarray}
    H_{0}&=&-t_x \sum_{l ,\sigma} \sum_{\langle i,j \rangle} 
\left(d^\dagger_{1,i,l,\sigma}d_{1,j,l,\sigma} +\text{H.c.}\right)     \label{eq:two-orbital}\\ 
    &&-t_z\sum_{l,\sigma }  \sum_{\langle i,j \rangle} 
\left(d^\dagger_{2,i,l,\sigma}d_{2,j,l,\sigma} +\text{H.c.}\right) \notag \\ 
    &&-t_{xz}\sum_{l,\sigma }\sum_{\langle i,j \rangle} 
    \left(( -1 )^{s_{ij}} d^\dagger_{1,i,l,\sigma }d_{2,j,l,\sigma } +\text{H.c.}\right) \notag \\ 
    &&-t^{\perp}_z \sum_{i,\sigma} 
\left(d^\dagger_{2,i,t,\sigma}d_{2,i,b,\sigma} +\text{H.c.}\right)
    +\Delta \sum_i (n_{i,1}-n_{i,2}) \notag .
\end{eqnarray}
Here, $l=t,b$ labels the layer index, and $\sigma=\uparrow,\downarrow$ denotes the spin. We denote $d_1$ and $d_2$ as the $d_{x^2-y^2}$ and $d_{z^2}$ orbitals, respectively. Interlayer hopping is included only for the $d_{2}$ orbital, reflecting its strong out-of-plane character, while direct interlayer hybridization of the $d_{1}$ orbital is negligible. The hopping parameters are estimated from density functional theory as $t_{x}=0.485$, $t_{z}=0.110$, $t_{xz}=0.239$, and $t^{\perp}_{z}=0.635$~\cite{luo2023bilayer}. The parameter $\Delta$ represents the crystal-field splitting between the two orbitals.
The factor $s_{ij}=1$ ($-1$) for bonds along the $x$ ($y$) direction. The corresponding bond-dependent sign factor $(-1)^{s_{ij}}$ encodes the symmetry of the $d_{x^2-y^2}$–$d_{z^2}$ hybridization on the square lattice.
On average, the electron filling is $n=2-x$ per site (summed over spin), with $x\approx0.5$ relevant to experiments, corresponding to orbital fillings $n_1 \approx 0.5$ and $n_2 \approx 1$. 
The resulting band structure hosts $\alpha$, $\beta$, and $\gamma$ Fermi pockets, as depicted in Fig.~\rev{1}(d). Previous studies have shown that the $\alpha$ and $\beta$ pockets exhibit mixed orbital character, while the $\gamma$ pocket is predominantly from the $d_{2}$ orbital. 

\subsection{B. Derivation of form factor in one-orbital model}
The effective interlayer
hopping between the $d_{x^2-y^2}$ orbitals is generated by a virtual process
through the $d_{z^2}$ orbital, as illustrated in Fig.\rev{1}(a) in Main-text:
$$
d_{x^2-y^2,t}
\rightarrow d_{z^2,t}
\rightarrow d_{z^2,b}
\rightarrow d_{x^2-y^2,b}.
$$
The intra-layer hybridization between $d_{x^2-y^2}$ and $d_{z^2}$ has the
form
$$
V_{xz}({\bf k}) \propto \cos k_x-\cos k_y,
$$
because the hopping amplitudes along the $x$ and $y$ directions have opposite
signs due to the $d_{x^2-y^2}$ orbital symmetry.
\\

Integrating out the intermediate $d_{z^2}$ states then gives
$$
t_{\perp}^{\rm eff}({\bf k})
\sim
\frac{
V_{xz}({\bf k}),t_{\perp}^{z},V_{zx}({\bf k})
}{\Delta^2}
\propto
(\cos k_x-\cos k_y)^2,
$$
where $t_{\perp}^{z}$ is the interlayer hopping between $d_{z^2}$ orbitals and
$\Delta$ is the energy separation to the virtual $d_{z^2}$ states. Therefore,
the effective one-orbital model contains
$$
\gamma({\bf k})=-t_\perp(\cos k_x-\cos k_y)^2.
$$

Moreover, this form factor is also important phenomenologically because it
captures the Fermi-surface structure of the original two-orbital model. Since
$(\cos k_x-\cos k_y)^2$ vanishes along the diagonal line
$(0,0)$--$(\pi,\pi)$, the effective interlayer hybridization disappears on this
line. As a result, the two $d_{x^2-y^2}$-dominated bands are not split along
the diagonal direction, consistent with the touching of the $\alpha$ and
$\beta$ pockets shown in two-band model, Fig.2(d).

\section{II. Details on Mean-field theory}
\label{sec:s2}
We consider the effective one orbital model, introduced in Eq.~(\textcolor{red}{1}),
\begin{eqnarray}
    H_{\mathrm{eff}}& =&
    \sum_{\bm{k},\sigma}
\phi_{\bm{k},\sigma}^
\dagger
   \left(\begin{array}{cc}
   \xi(\bm{k})+D & \gamma(\bm{k})\\
   \gamma_(\bm{k}) &    \xi(\bm{k})-D
    \end{array}
    \right)
\phi_{\bm{k},\sigma}
    =
      \sum_{\bm{k}}
\phi_{\bm{k},\sigma}^\dagger \mathcal{H}_N (\bm{k})
      \phi_{\bm{k},\sigma}
    \end{eqnarray}
    with  $\phi_{\bm{k},\sigma}^T=(c_{t,k,\sigma},c_{b,k,\sigma})$ and
\begin{eqnarray}
\xi({\bm{k}})&=&-t(\cos k_x + \cos k_y)-\mu,\\
\gamma(\bm{k})&=&-t_\perp (\cos k_x -\cos k_y)^2,
\end{eqnarray}
and $D$ is displacement field strength. Note that we implicitly set $\mu$ to fix the electron density $n=1-x$.

We next incorporate the phenomenological attractive interaction as similar setup of BCS theory. Specifically interlayer density interaction $J_\perp>0$ is considered,
    \begin{eqnarray}
 H_\mathrm{int}&=& \frac{J_\perp}{2} \sum_{i} 
    \left[
    S_{t,i}\cdot S_{b,i}-\frac{1}{4}
    n_{t,i} n_{b,i}\right].
\end{eqnarray}
We then employ a standard mean-field analysis with an ansatz for the interlayer pairing order parameter,
\begin{eqnarray}
    \langle c_{t,i \uparrow}c_{b,i \downarrow}\rangle
    =-\langle c_{t,i \downarrow}c_{b,i \uparrow}\rangle
    =\Delta_{\bm{Q}} \exp(i \bm{R}_i \cdot \bm{Q}),
    \label{eq:pdw_order}
\end{eqnarray}
where $\bm{Q}$ denotes the center-of-mass momentum of the Cooper pair.

The mean-field Hamiltonian in Nambu basis $\Phi_{\bm{k}} =         \left( 
    \begin{array}{cc}
      \phi_{\bm k^+,\uparrow} 
        \phi_{-\bm k^-,\downarrow}^*
    \end{array}
    \right)$ is 
\begin{eqnarray}
    H_\mathrm{MF}
    &=&
    \frac{1}{2}
    \sum_{\bm k\in B.Z.}
\Phi_{\bm{k}}^\dagger
\mathcal{H}_\mathrm{BdG}(\bm{k})
\Phi_{\bm{k}} +E_0\nonumber
    \end{eqnarray}
with 
    \begin{eqnarray}
  \mathcal{H}_\mathrm{BdG}&=&  \left( 
    \begin{array}{cc}
       \mathcal{H}_N(\bm k^+)  &  \mathcal{H}_\mathrm{pair} \\
\mathcal{H}_\mathrm{pair}^\dagger  & 
      -\mathcal{H}_N^T(-\bm k^-)
    \end{array}
    \right)
    ,
    \quad
    \mathcal{H}_\mathrm{pair}=-J_\perp\Delta
      \left( 
    \begin{array}{cc}
0 & 1\\
1 & 0 
    \end{array}
    \right), 
\end{eqnarray}
and constant energy term 
\begin{eqnarray}
        E_0 = \frac{1}{2}\sum_{\bm k \in B.Z.}\mathrm{Tr}
\left( \mathcal{H}_N(-\bm k^-)\right)
  +NJ_\perp|\Delta|^2. 
\end{eqnarray}
Here, $\bm{k}^{\pm}=\bm{k}\pm \bm{Q}/2$ is used.

For each $\bm{Q}$, the pairing amplitude $\Delta_{\bm{Q}}$ is determined self-consistently from $H_{\mathrm{MF}}$ and Eq.~\ref{eq:pdw_order}. 
To further identify the phase among all $\bm{Q}$, we should compute the mean-field free-energy,
\begin{eqnarray}
    F_\mathrm{MF}
    &=& -T \log \left(\mathrm{Tr}
    \left(e^{-{H_\mathrm{MF}/T}} \right)
    \right)
    + E_0, 
\end{eqnarray}
and compare $F$ for different $\bm{Q}$. The above mean-field analysis allows us to determine the energetically favored state as a function of system parameters.

\subsection{A. Mean-field phase diagram without form factor}
\begin{figure}[h]
\centering\includegraphics[width=0.6\linewidth]{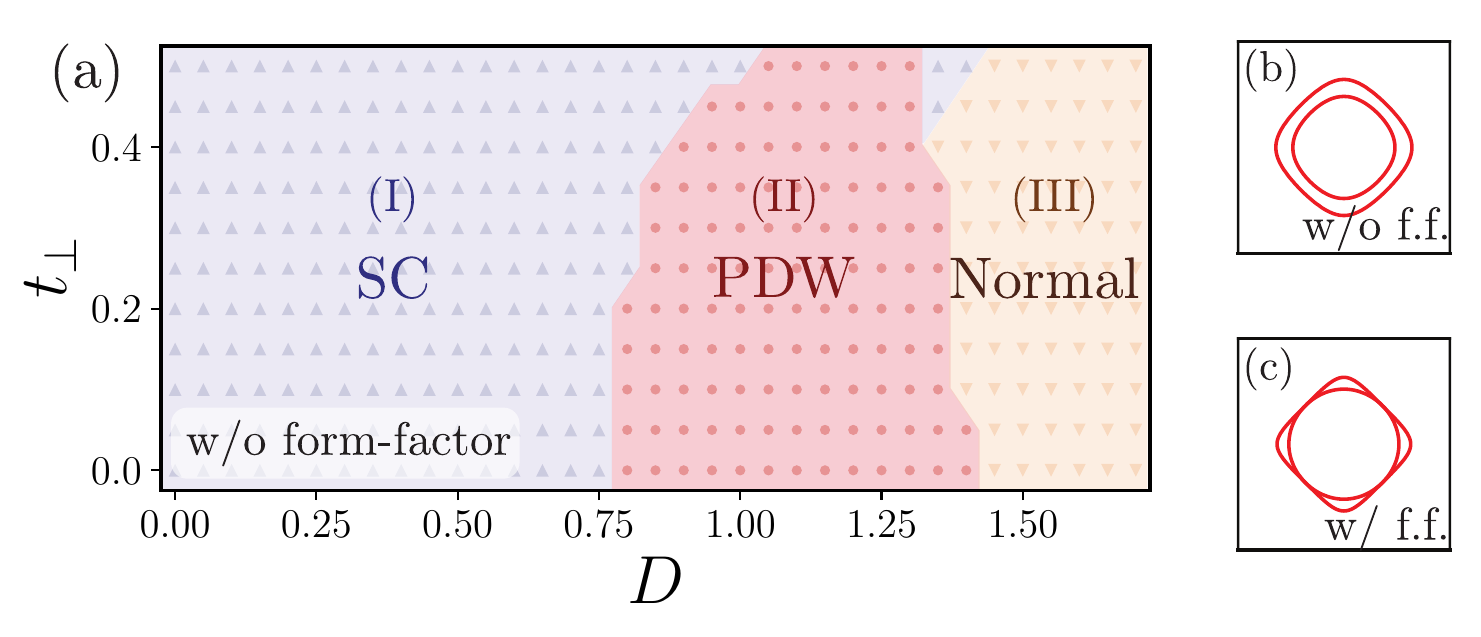}
    \caption{(a) Zero-temperature mean-field phase diagram without a form factor, $t_\perp(\bm{k}) = t_\perp$.
We fix $J_\perp = 2$ and $x = 1/2$.
(b,c) Fermi surfaces without (b) and with (c) the form factor at $D=0$.
In the absence of the form factor, the nesting vector is relatively well defined, which stabilizes the PDW phase, compared to Fig.~\rev{3}(a).
}
    \label{fig:s1}
\end{figure}

We also present the zero-temperature mean-field phase diagram without the momentum-dependent form factor, taking $\gamma(\bm{k}) = t_\perp$.
The resulting phase diagram exhibits the same qualitative features as that obtained with the form factor shown in Fig.~3(a) of main-text. As illustrated in Fig.~\ref{fig:s1}, the results  show that PDW is more stable than without form factor case. This is related to the shape of Fermi-surface, where the nesting vector is more well-defined in the absence of the form factor. 



\section{III. Details on computing $\alpha_\mathrm{dis}$}\label{sec:s3}

\subsection{A. Setup}
\label{sec:s3a}
We again consider the two-band normal-state Hamiltonian,
\begin{eqnarray}
H_N = \xi(\bm{k}) \tau_0 + t_\perp(\bm{k}) \tau_x,
\end{eqnarray}
where $\tau_i$ ($i=0,1,2,3$) are Pauli matrices defined in the layer basis.
The corresponding band dispersions are given by
$\xi_\pm(\bm{k}) = \xi(\bm{k}) \pm t_\perp(\bm{k})$,
which become degenerate in the absence of interlayer hopping, $t_\perp = 0$.

Including interlayer pairing, the Bogoliubov--de Gennes (BdG) Hamiltonian can be written as
\begin{eqnarray}
H_{\mathrm{BdG}} =
\left(
\begin{array}{cc}
H_N(\bm{k}^+) & \Delta \\
\Delta^\dagger & -H_N^T(-\bm{k}^-)
\end{array}
\right),
\end{eqnarray}
with $\Delta = \Delta_0 \, \tau_1$.

We consider a general spin-independent disorder potential in the two-layer basis,
\begin{eqnarray}
V_{\mathrm{dis}}
=
\sum_{i=0,1,2,3} V_i \, \tau_i .
\label{eq:v_dis}
\end{eqnarray}
Here we restrict our analysis to the $s$-wave scattering channel, which is momentum independent.

It is convenient to transform from the layer basis to the band basis via a unitary transformation.
Under this transformation, the Pauli matrices in the layer basis are mapped to those in the band basis as
\begin{eqnarray}
\left\{
\tau_0, \tau_1, \tau_2, \tau_3
\right\}
\rightarrow
\left\{
\tilde{\tau}_0, \tilde{\tau}_3, \tilde{\tau}_2, -\tilde{\tau}_1
\right\}.
\label{eq:pauli}
\end{eqnarray}

\subsection{B. 1st order Born approximation}
\label{sec:s3b}
We then employ the conventional first-order Born approximation (1BA), treating $V_{\mathrm{dis}}$ as a perturbation~\cite{abrikosov1961problem,bruus2004many}.
The corresponding self-energy is given by
\begin{eqnarray}
\Sigma(i \omega_n)
=
n_{\mathrm{imp}}
\int_{\bm{k}}
\tilde{V}_{\mathrm{dis}}
\, G_{\mathrm{BdG}}(\bm{k}, i \omega_n)\,
\tilde{V}_{\mathrm{dis}},
\label{eq:self_1BA}
\end{eqnarray}
where $G_{\mathrm{BdG}}(\bm{k}, i \omega_n)=(i\omega_n-H_\mathrm{BdG})^{-1}$ is BdG Green's function and $\tilde{V}_{\mathrm{dis}}$ denotes the disorder potential written in the Nambu basis. $i \omega_n$ is the fermionic Matsubara frequency. A diagrammatic representation is shown in Fig.~\ref{fig:s1}(a).

We then decompose the self-energy in Eq.~\ref{eq:self_1BA} into contributions that renormalize the frequency and the pairing amplitude, denoted by $\Sigma_\omega$ and $\Sigma_\Delta$, respectively.
These are given by
\begin{eqnarray}
\Sigma_{\omega}(\bm{k}, i\omega_n)
=
\lim_{\Delta_{0}\rightarrow 0}
\int_{\bm{k}}
\frac{n_{\mathrm{imp}} V^2}{2}\,
\mathrm{Tr}
\Big[
\hat{V}\,
G(\bm{k}, i \omega_n)\,
\hat{V}
\Big],
\label{eq:self_omega}
\end{eqnarray}
and
\begin{eqnarray}
\frac{\Sigma_{\Delta}(\bm{k}, i \omega_n)}{\Delta_{0}}
=
-\lim_{\Delta_{0}\rightarrow 0}
\int_{\bm{k}}
\frac{n_{\mathrm{imp}} V^2}{2\Delta_{0}}\,
\mathrm{Tr}
\Big[
\hat{V}\,
F(\bm{k}, i  \omega_n)\,
\hat{V}^{*}\,
\hat{\Delta}^{\dagger}
\Big],\label{eq:self_pair}
\end{eqnarray}
where $G$ and $F$ denote the diagonal and off-diagonal blocks of the BdG Green’s function $\mathcal{G}_{\mathrm{BdG}}$, respectively.
The limit $\Delta_{0}\rightarrow 0$ is taken since we will focus on the critical temperature, assuming that the superconducting transition is a continuous second-order phase transition. The hat notation in disorder potential and pairing denotes the normalized matrix, i.e. $\hat{\Delta}=\tau_1$.

\begin{figure}[h]
    \centering
\includegraphics[width=0.5\linewidth]{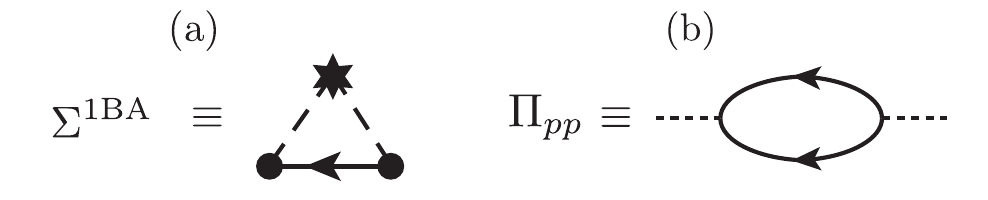}
    \caption{Diagrammatic representation of (a) the first-order Born approximation and (b) the particle–particle susceptibility.
}
    \label{fig:s2}
\end{figure}

\subsection{C. $t_\perp = 0$ case}
\label{sec:s3c}

We begin with the simple case $t_\perp = 0$, where the two bands are degenerate with dispersion $\xi_{\bm{k}}$.
In the limit $\Delta_0 \rightarrow 0$, the normal and anomalous Green’s functions simplify to
\begin{eqnarray}
\lim_{\Delta_{0}\rightarrow 0} G(\bm{k}, i\omega_n)
&=&
-\frac{i\omega_n + \xi_{\bm{k}}}{\omega_n^2 + \xi_{\bm{k}}^{2}},
\quad
\lim_{\Delta_{0}\rightarrow 0}
\frac{F(\bm{k}, i\omega_n)}{\Delta_{0}}
=
-\frac{\hat{\Delta}}{\omega_n^2 + \xi_{\bm{k}}^{2}} .
\label{eq:green_no_tp}
\end{eqnarray}

Substituting these expressions into Eqs.~\ref{eq:self_omega} and \ref{eq:self_pair}, we obtain
\begin{eqnarray}
\Sigma_{\omega}(\bm{k}, i\omega_n)
&=&
-\frac{i\omega_n}{|\omega_n|}
\pi N(0)n_{\mathrm{imp}} V^2 ,
\label{eq:sigma_omega_notp}
\end{eqnarray}
and
\begin{eqnarray}
\Sigma_{\Delta}(\bm{k}, i \omega_n)
&=&
\frac{\Delta_{0}}{|\omega_n|}
\pi N(0)\frac{n_{\mathrm{imp}} V^2}{2}\,
\mathrm{Tr}\Big[
\hat{V}\,
\hat{\Delta}\,
\hat{V}^{*}\,
\hat{\Delta}^{\dagger}
\Big],
\label{eq:sigma_delta_notp}
\end{eqnarray}
where we used $\int_{\bm{k}} = N(0)\int d\xi$, with $N(0)$ the density of states at the Fermi energy.

The trace part in Eq.~\ref{eq:sigma_delta_notp} can be simplified using
\begin{eqnarray}
\mathrm{Tr}
\Big[
\hat{V}\,
\hat{\Delta}\,
\hat{V}^{*}\,
\hat{\Delta}^{\dagger}
\Big]
=
2-\frac{1}{2}
\mathrm{Tr}
\Big[
\hat{F}_{c}^{\dagger}
\hat{F}_{c}
\Big],
\label{eq:equality}
\end{eqnarray}
where the superconducting fitness function is defined as
\begin{eqnarray}
F_c = V\Delta - \Delta V^* .
\end{eqnarray}

Substituting Eq.~\ref{eq:equality} into Eq.~\ref{eq:sigma_delta_notp}, we obtain
\begin{eqnarray}
\Sigma_{\Delta}(\bm{k}, i \omega_n)
=
\frac{\Delta_{0}}{|\omega_n|}
\pi N(0)n_{\mathrm{imp}} V^2
\left[
1-\frac{1}{4}
\mathrm{Tr}
\Big(
\hat{F}_{c}^{\dagger}
\hat{F}_{c}
\Big)
\right].
\end{eqnarray}

The renormalized Matsubara frequency and pairing amplitude are therefore given by
\begin{eqnarray}
\tilde{\omega}_n
&=&
\omega_n
+
\frac{\tilde{\omega}_n}{|\tilde{\omega}_n|}
\Gamma,
\label{eq:tilde_omega_no_tp}
\\
\tilde{\Delta}_{0}
&=&
\Delta_{0}
+
\frac{\tilde{\Delta}_{0}}{|\tilde{\omega}_n|}
\left[
1-\frac{1}{4}\mathrm{Tr}
\Big(
\tilde{F}_c^\dagger
\tilde{F}_c
\Big)
\right]
\Gamma,
\label{eq:tilde_delta_no_tp}
\end{eqnarray}
where scattering frequency, $\Gamma\equiv \pi N(0)n_{\mathrm{imp}} V^2 =\tau^{-1}$ is used.  

As we show in the next subsection, the ratio between $\tilde{\omega}_n / \tilde{\Delta}_0\equiv u_n$ is central quantity in determining the disorder-induced suppression of the critical temperature, as obtained from the linearized gap equation.
From Eqs.~\ref{eq:tilde_omega_no_tp}-~\ref{eq:tilde_delta_no_tp}, one can find 
\begin{eqnarray}
    \frac{\omega_n}{\Delta_0}=u_n
\left[ 1 - \frac{\alpha_{\mathrm{dis}}}{\Delta_0} \frac{\Gamma}{|u_n|} \right]
    \label{eq:u_no_tp}
\end{eqnarray}
with defining depairing coefficient
\begin{eqnarray}
\alpha_{\mathrm{dis}}
=\frac{1}{4}
\mathrm{Tr}
\Big(
\tilde{F}_c^\dagger
\tilde{F}_c
\Big).
\end{eqnarray}
If the condition $\tilde{\omega}_n / \tilde{\Delta}_0=\omega_n/\Delta_0$ is satisfied, meaning that the frequency and pairing renormalizations are identical (equivalently, $\alpha_{\mathrm{dis}}=0$), the superconducting state is robust against the corresponding disorder channel.

Finally, for the disorder channels $\hat{V}=\tau_i$ defined in Eq.~\ref{eq:v_dis} and interlayer pairing
$\hat{\Delta}=\tau_1$, we find
\begin{eqnarray}
F_c = 0,
\quad
\text{for all channels except } \tau_3 .
\label{eq:criteria}
\end{eqnarray}
consequently, $\alpha_\mathrm{dis}=0$ for all channels except $\tau_3$, while $\alpha_\mathrm{dis}=2$ for $\tau_3$. 
This result directly leads to Table~\rev{I}.
The $\tau_3$ disorder channel breaks the composite $\mathcal{M}\mathcal{T}$ symmetry, while all other channels preserve it.

\subsubsection{Renormalized gap equation and $T_c$}

We now evaluate the critical temperature in the presence (absence) of disorder, denoted by $T_c$ ($T_{c,0}$).
In the presence of impurity scattering, the gap equation is modified as
\begin{eqnarray}
\frac{1}{2g}
=
\frac{T_{c,0}}{\mathcal{V}}
\sum_{\bm{k},\omega_n}
\frac{1}{\omega_n^{2}+\xi_{\bm{k}}^{2}}
\;\rightarrow\;
\frac{\Delta_0}{2g}
=
\frac{T_{c}}{\mathcal{V}}
\sum_{\bm{k},\omega_n}
\frac{\tilde{\Delta}_{0}}{\tilde{\omega}_n^{2}+\xi_{\bm{k}}^{2}} ,
\end{eqnarray}
where $g>0$ denotes the attractive pairing interaction. The gap equation is achieved by the second order Ginzburg Landau diagram as illustrated in Fig.~\ref{fig:s2}(b).

Replacing the momentum summation by an integral over energies near the Fermi surface and using Eqs.~\ref{eq:tilde_omega_no_tp}-\ref{eq:tilde_delta_no_tp}, the gap equation can be rewritten as 
\begin{eqnarray}
\frac{1}{2g}
\simeq
\pi N(0) T_{c,0}
\!\!\sum_{|\omega_n|<\Lambda_{UV}}\!\!
\frac{1}{|\omega_n|}
\;\rightarrow\;
\frac{\Delta_0}{2g}
\simeq
\pi N(0) T_{c}
\!\!\sum_{|\omega_n|<\Lambda_{UV}}\!\!
\frac{\tilde{\Delta}_{0}}{|\tilde{\omega}_n|},
\end{eqnarray}
where $\Lambda_{UV}$ is an ultraviolet energy cutoff imposed on the Matsubara frequencies.

Substituting the renormalized parameters from Eq.~\ref{eq:u_no_tp}, we obtain
\begin{eqnarray}
\frac{1}{2g}
\simeq 
 \pi N(0) T_{c}
\sum_{|\omega_n|<\Lambda_{UV}}
\Bigg[
\frac{1}{\omega_n }
-\frac{\alpha_\mathrm{dis}}{\omega_n^2}\Gamma
\Bigg]
\label{se56}
\end{eqnarray}

To perform the Matsubara summation in Eq.~\ref{se56}, we make use of the digamma function $\Psi(z)$ and its derivative $\Psi^{(1)}(z)=d\Psi(z)/dz$ defined as
\begin{eqnarray}
\Psi(z)
&=&
-\xi
-
\sum_{n=0}^{\infty}
\Bigg[
\frac{1}{n+z}
-
\frac{1}{n+1}
\Bigg],
\quad
\Psi^{(1)}(z)
=
\sum_{n=0}^{\infty}
\frac{1}{(n+z)^2},
\nonumber
\end{eqnarray}
where $\xi$ denotes the Euler constant.
For large $|z|$, the asymptotic forms
$\Psi(z)\simeq \log z$ and $\Psi^{(1)}(z)\simeq 1/z$ can be used.

Applying these identities yields
\begin{eqnarray}
2\pi T_{c}
\sum_{\omega_n=0}^{\Lambda_{UV}}
\frac{1}{\omega_n}
&=&
-
\Bigg[
\Psi\!\left( \frac{1}{2}\right)
-
\Psi\!\left(
\frac{1}{2}
+
\frac{\Lambda_{UV}}{2\pi T_{c}}
\right)
\Bigg]
\nonumber\\
&\simeq&
-
\Bigg[
\Psi\!\left( \frac{1}{2}\right)
-
\log\!\left(
\frac{\Lambda_{UV}}{2\pi T_c}
\right)
\Bigg],
\end{eqnarray}
and
\begin{eqnarray}
2\pi T_{c}
\sum_{\omega_n=0}^{\Lambda_{UV}}
\frac{1}{\omega_n^2}
&=&
\frac{1}{2\pi T_c}
\Bigg[
\Psi^{(1)}\!\left( \frac{1}{2}\right)
-
\Psi^{(1)}\!\left(
\frac{1}{2}
+
\frac{\Lambda_{UV}}{2\pi T_{c}}
\right)
\Bigg]
\nonumber\\
&\simeq&
\frac{1}{2\pi T_c}
\Bigg[
\frac{\pi^2}{2}
-
\frac{2\pi T_{c}}{\Lambda_{UV}}
\Bigg],
\end{eqnarray}
valid in the limit $T_c \ll \Lambda_{UV}$.

Finally, we obtain the disorder-suppressed critical temperature,
\begin{eqnarray}
\log\left(\frac{T_{c}}{T_{c,0}}\right)
&\simeq &-\frac{\pi}{4}
\frac{\alpha_{\mathrm{dis}}}{T_{c,0}}
\Gamma,
\label{eq:Tc_disorder}
\end{eqnarray}
where the parameter $\alpha_{\mathrm{dis}}$ quantifies the suppression of the critical temperature due to disorder.
\begin{figure}[h]
    \centering
    \includegraphics[width=0.55\linewidth]{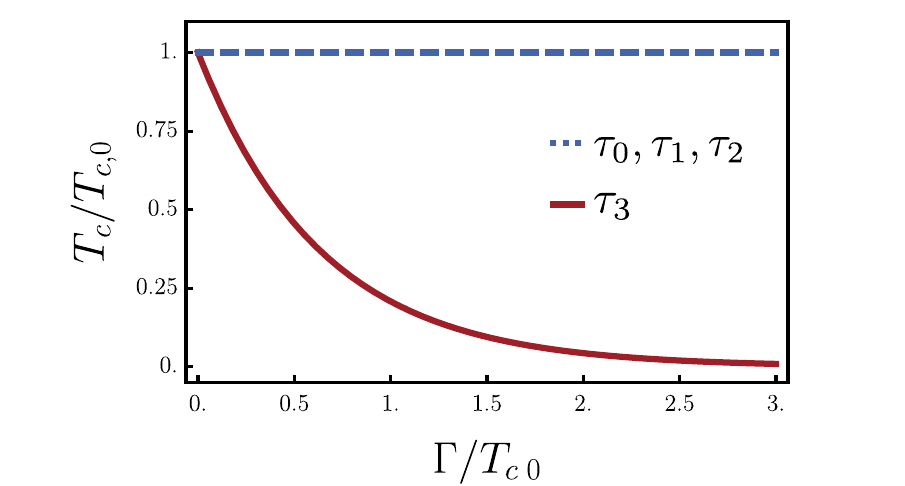}
    \caption{
Transition temperature $T_c$ as a function of the scattering rate $\Gamma$, obtained from Eq.~\ref{eq:Tc_disorder}.
The $\tau_3$ disorder channel, which breaks mirror symmetry, suppresses superconductivity (i.e. $\alpha_\mathrm{dis}=2$) irrespective of whether $t_\perp=0$ or finite.}
    \label{fig:s3}
\end{figure}

\subsection{D. $t_\perp \neq 0$ case}\label{sec:s3d}
For the case where $t_\perp \neq 0$, the band splitting modifies the dispersion to $\xi_\pm(\bm{k}) = \xi(\bm{k}) \pm |t_\perp|$, which may modify the expression for the depairing coefficient $\alpha_{\mathrm{dis}}$. Following the framework introduced in Sec.~III~A, we evaluate the normal and anomalous Green’s functions, $G$ and $F$, in the limit $\Delta_{0} \rightarrow 0$ to calculate the self-energies.

In the band basis, the normal Green’s function is expressed as,
\begin{eqnarray}
G_N(\bm{k}, i\omega_n)
&=&
\frac{1}{2}\left[(i\omega_n - \xi_+(\bm k))^{-1} + (i\omega_n - \xi_-(\bm k))^{-1}\right] \tilde{\tau}_0 \nonumber\\
&+&
\frac{1}{2}\left[(i\omega_n - \xi_+(\bm k))^{-1} - (i\omega_n - \xi_-(\bm k))^{-1}\right] \tilde{\tau}_3,
\label{eq:green_tp1}
\end{eqnarray}
where the tilde notation denotes Pauli matrices defined in the band basis. The anomalous Green’s function in the same limit is given by,
\begin{eqnarray}
\lim_{\Delta_{0}\rightarrow 0} \frac{F(\bm{k}, i\omega_n)}{\Delta_{0}}
&=& G_N \hat{\Delta} G_N^{*} = G_N G_N^{*} \hat{\Delta},
\label{eq:green_tp2}
\end{eqnarray}
where we have used the fact that $[G_N, \hat{\Delta}] = 0$, as both operators are diagonal in the band basis (i.e., expressed in $\tilde{\tau}_0$ and $\tilde{\tau}_3$).

Substituting these expressions into the self-energy equations, we obtain the results for $t_\perp \neq 0$. For the frequency self-energy, we find,
\begin{eqnarray}
\lim_{\Delta_{0}\rightarrow 0} \mathrm{Tr}\Big[ \hat{V} G(\bm{k}, i\omega_n) \hat{V} \Big]
&=& \mathrm{Tr}\Big[ \left( G_N \hat{V} + [\hat{V}, G_N] \right) \hat{V} \Big] = \mathrm{Tr}\Big[ G_N \hat{V} \hat{V} \Big] \label{eq:s36} \\
&=& -\left( \frac{i\omega_n+\xi_+}{\omega_n^2 +\xi_+^2} + \frac{i\omega_n+\xi_-}{\omega_n^2 +\xi_-^2} \right), \label{eq:G_tp}
\end{eqnarray}
and for the pairing self-energy,
\begin{eqnarray}
\lim_{\Delta_{0}\rightarrow 0} \frac{1}{\Delta_0} \mathrm{Tr} \Big[ \hat{V} F(\bm{k}, i\omega_n) \hat{V}^{*} \hat{\Delta}^{\dagger} \Big]
&=& \mathrm{Tr} \Big[ \big( [\hat{V}, G_N G_N^{*}] + G_N G_N^{*}\hat{V} \big) \hat{\Delta} \hat{V}^{*} \hat{\Delta}^{\dagger} \Big] \nonumber \\
&=& \mathrm{Tr} \Big[ G_N G_N^{*} \hat{V} \hat{\Delta} \hat{V}^{*} \hat{\Delta}^{\dagger} \Big] \label{eq:s38} \\
&=& -\left( \frac{1}{\omega_n^2 +\xi_+^2} + \frac{1}{\omega_n^2 +\xi_-^2} \right) \times \Big[ 1-\frac{1}{4}\mathrm{Tr} (\hat{F}_c^\dagger \hat{F}_c) \Big]. \label{eq:F_tp}
\end{eqnarray}
In Eqs.~\eqref{eq:s36} and \eqref{eq:s38}, the terms involving commutators vanish upon taking the trace $\mathrm{Tr}([\hat{V}, G_N]\hat{V}) = \mathrm{Tr}([\hat{V}, G_N G_N^*]\hat{\Delta} \hat{V}^* \hat{\Delta}^\dagger) =0$, because $\hat{V}$ is one of Pauli matrix $\tilde{\tau}_i$, and $G_N$ and $G_N G_N^*$ are expressed only by $\{\tilde{\tau}_0, \tilde{\tau}_3\}$.

Performing the momentum summation, the effective density of states (DOS) is modified to $\bar{N}(0) = \frac{1}{2}(N_+(0) + N_-(0))$. Crucially, however, the ratio $u_n = \tilde{\omega}_n / \tilde{\Delta}_n$ remains identical to the $t_\perp = 0$ case (See Eq.~\ref{eq:u_no_tp}),
\begin{eqnarray}
u_n \equiv \frac{\omega_n}{\Delta_0} = u_n \left[ 1 - \frac{\alpha_{\mathrm{dis}}}{\Delta_0} \frac{\Gamma}{|u_n|} \right], \label{eq:u}
\end{eqnarray}
with the same depairing coefficient,
\begin{eqnarray}
\alpha_{\mathrm{dis}} = \frac{1}{4} \mathrm{Tr} (\tilde{F}_c^\dagger \tilde{F}_c).
\end{eqnarray}

\subsubsection{Renormalized gap equation and $T_c$}
We now evaluate the critical temperature in the presence of disorder, $T_c$. The linearized gap equation is given by,
\begin{eqnarray}
\frac{1}{g} = \frac{T_{c}}{\mathcal{V}} \sum_{\bm{k},\omega_n} \left[ \frac{\tilde{\Delta}_{0}}{\tilde{\omega}_n^{2}+\xi_{+,\bm{k}}^{2}} + \frac{\tilde{\Delta}_{0}}{\tilde{\omega}_n^{2}+\xi_{-,\bm{k}}^{2}} \right].
\end{eqnarray}

After performing the summation over momentum and Matsubara frequencies, we observe that since the ratio $u_n=\tilde \omega_n / \tilde \Delta_0$ is derived to be unchanged, the functional form of the $T_c$ suppression remains the same with the $t_\perp = 0$ case,
\begin{eqnarray}
\log\left(\frac{T_{c}}{T_{c,0}}\right) = -\frac{\pi}{4} \frac{\alpha_{\mathrm{dis}}}{T_{c,0}} \Gamma. \label{eq:Tc_disorder}
\end{eqnarray}

While the same ratio between $T_{c}/T_{c,0}$ holds, the clean-limit critical temperature $T_{c,0}$ itself is rescaled by the change in the DOS,
\begin{eqnarray}
\frac{T_{c,0}(t_\perp)}{T_{c,0}(t_\perp=0)} = \frac{2N(0)}{N_+(0) + N_-(0)},
\end{eqnarray}
where $N_\pm(0)$ denotes the DOS of the $\xi_\pm$ band at the Fermi level.

\end{document}